\documentclass{article}
\usepackage{amssymb}
\usepackage{graphicx}
\usepackage{amsmath}

\input{tcilatex}

\begin{document}

\begin{center}
{\LARGE Astrophysical and Cosmological Tests of Quantum Theory}

\bigskip\bigskip

\bigskip Antony Valentini\footnote{%
Present address: Centre de Physique Th\'{e}orique, Campus de Luminy, Case
907, 13288 Marseille cedex 9, France. Email: valentini@cpt.univ-mrs.fr.}

\bigskip

\bigskip

\textit{Perimeter Institute for Theoretical Physics,}

\textit{31 Caroline Street North, Waterloo, Ontario N2L 2Y5, Canada.}
\end{center}

\bigskip

\bigskip

\bigskip

We discuss several proposals for astrophysical and cosmological tests of
quantum theory. The tests are motivated by deterministic hidden-variables
theories, and in particular by the view that quantum physics is merely an
effective theory of an equilibrium state. The proposed tests involve
searching for nonequilibrium violations of quantum theory in: primordial
inflaton fluctuations imprinted on the cosmic microwave background, relic
cosmological particles, Hawking radiation, photons with entangled partners
inside black holes, neutrino oscillations, and particles from very distant
sources.

\bigskip

\bigskip

\bigskip

1 Introduction

2 Quantum Nonequilibrium: What, When and Where?

3 Inflation as a Test of Quantum Theory in the Early Universe

4 Relic Nonequilibrium Particles

5 Tests of Quantum Theory with Black Holes

6 Neutrino Oscillations

7 Particles from Very Distant Sources

8 Conclusion

\bigskip

\bigskip

\bigskip

Published in: J. Phys. A: Math. Theor. \textbf{40}, 3285--3303 (2007).
(Special issue: \textit{The Quantum Universe}, eds. S. Adler, A. Bassi, F.
Dowker and D. D\"{u}rr, dedicated to Prof. G.-C. Ghirardi on the occasion of
his seventieth birthday.)

\bigskip

\bigskip

\bigskip

\bigskip

\bigskip

\bigskip

\bigskip

\bigskip

\bigskip

\bigskip

\bigskip

\bigskip

\bigskip

\bigskip

\bigskip

\bigskip

\bigskip

\bigskip

\bigskip

\bigskip

\bigskip

\bigskip

\bigskip

\bigskip

\bigskip

\bigskip

\bigskip

\bigskip

\bigskip

\bigskip

\bigskip

\bigskip

\bigskip

\bigskip

\bigskip

\bigskip

\bigskip

\bigskip

\bigskip

\bigskip

\bigskip

\bigskip

\bigskip

\bigskip\bigskip

\section{Introduction}

It is important that we continue to test quantum theory in new and extreme
conditions: as with any scientific theory, its domain of validity can be
determined only by experiment. For this purpose, it is helpful to have
theories that agree with quantum theory in some limit, and deviate from it
outside that limit. Examples of such theories include models of wave
function collapse, pioneered by Pearle [1--3] and by Ghirardi, Rimini and
Weber \cite{GRW86}, and hidden-variables theories with nonstandard
probability distributions (`quantum nonequilibrium'), advocated in
particular by the author [5--13].

While it is possible that quantum theory might turn out to break down in a
completely unexpected way, and in a completely unexpected place, the chances
of a successful detection of a breakdown would seem higher, the better
motivated the theory describing the breakdown.

For some 25 years, extreme tests of quantum theory focussed mostly on
experiments demonstrating violations of Bell's inequality. These tests were
well motivated: at the time (say in the 1970s), it was reasonable to suspect
that locality might force a deviation from quantum correlations for
entangled states of widely-separated systems. However, as the evidence for
violations of Bell's inequalities accumulated, the long-range correlations
predicted by quantum theory came to be widely accepted as a fact of nature,
and the known domain of validity of quantum theory was extended into an
important region.

Tests of collapse models again stem from a compelling motivation: to test
the superposition of quantum states as far as possible into the macroscopic
regime. Will a sufficiently macroscopic superposition decay via corrections
to the Schr\"{o}dinger equation? Such experiments are still being carried
out, and again (for as long as they prove negative) extend our confidence in
the validity of quantum theory in an important way.

In this paper, we discuss a number of new proposals for extreme tests of
quantum theory, proposals that are motivated by thinking about quantum
physics from the point of view of deterministic hidden variables.

A deterministic hidden-variables theory provides a mapping $\omega
=\omega(M,\lambda)$ from initial (`hidden') parameters $\lambda$ to outcomes 
$\omega$ of a quantum experiment (or `measurement') specified by the
settings $M$ of macroscopic equipment. In addition, in order to make contact
with the statistics observed over an ensemble of similar experiments (with
fixed $M$ and variable $\lambda$), it must be assumed that over an ensemble
the hidden variables $\lambda$ have some distribution $\rho(\lambda)$, so
that (for example) the expectation value of $\omega$ will be given by%
\begin{equation}
\left\langle \omega\right\rangle =\int d\lambda\ \rho(\lambda)\omega
(M,\lambda)\ .
\end{equation}
For the hidden-variables theory to provide a successful account of quantum
phenomena, there must exist a particular distribution $\rho_{\mathrm{QT}%
}(\lambda)$ such that all corresponding expectation values $\left\langle
\omega\right\rangle _{\mathrm{QT}}$ match the prediction $\left\langle
\omega\right\rangle _{\mathrm{QT}}=\mathrm{Tr}(\hat{\rho}\hat{\Omega})$ of
standard quantum theory (for some density operator $\hat{\rho}$ and
`observable' $\hat{\Omega}$).

A concrete example is provided by the pilot-wave theory of de Broglie \cite%
{deB28} and Bohm \cite{B52}.\footnote{%
At the 1927 Solvay conference, de Broglie proposed what we now know as the
first-order pilot-wave dynamics of a (nonrelativistic) many-body system,
with a guiding wave in configuration space determining the particle
velocities, and he applied it to simple quantum phenomena such as
interference, diffraction, and atomic transitions. In 1952, Bohm showed that
the general quantum theory of measurement was a consequence of de Broglie's
dynamics (when applied to an initial equilibrium ensemble), even though Bohm
actually wrote the dynamics in a pseudo-Newtonian or second-order form based
on acceleration. For a detailed analysis of de Broglie's construction of
pilot-wave theory, as well as for a full discussion of the respective
contributions of de Broglie and Bohm, see ref. \cite{BV06} (which also
includes an English translation of de Broglie's 1927 Solvay report).} There,
the outcome of a single run of an experiment is determined by the initial
(`hidden') configuration $X(0)$ of the system, together with the initial
guiding wave function $\Psi (X,0)$, so that $\lambda $ consists of the pair $%
X(0)$, $\Psi (X,0)$. For an ensemble with the same $\Psi (X,0)$ (and the
same apparatus settings $M$), we have $\lambda =X(0)$, and the quantum
equilibrium distribution $\rho _{\mathrm{QT}}(\lambda )$ is given by $P_{%
\mathrm{QT}}(X,0)=\left\vert \Psi (X,0)\right\vert ^{2}$.

It is not usually appreciated that the distribution $\rho _{\mathrm{QT}%
}(\lambda )$ is conceptually quite distinct from the mapping $\omega =\omega
(M,\lambda )$. The latter is a property of each individual run of the
experiment, specifying the `dynamics' whereby each value of $\lambda $
determines an outcome $\omega $; while the former is a property of the
ensemble, specifying the distribution of `initial conditions' for the
parameters $\lambda $. As we have argued at length elsewhere [5--13], if one
takes deterministic hidden-variables theories seriously, one must conclude
that quantum theory is merely the phenomenology of a special `quantum
equilibrium' distribution $\rho _{\mathrm{QT}}(\lambda )$. In principle,
there exists a wider physics beyond the domain of quantum theory, with
`nonequilibrium' distributions $\rho (\lambda )\neq \rho _{\mathrm{QT}%
}(\lambda )$ and non-quantum expectation values $\left\langle \omega
\right\rangle \neq \left\langle \omega \right\rangle _{\mathrm{QT}}$. This
paper concerns the possibility of detecting such deviations from quantum
theory, through astrophysical and cosmological observations.

\section{Quantum Nonequilibrium: What, When and Where?}

What exactly should one look for? Quantum nonequilibrium opens up an immense
range of possible new phenomena. Here, we focus on deviations from the
following quintessentially quantum effects:

\begin{itemize}
\item Single-particle interference. For example, in a double-slit experiment
with particles of wavelength $2\pi/k$, incident on a screen with slits
separated by a distance $a$, at large distances behind the screen quantum
theory predicts a modulation%
\begin{equation}
\left\vert \psi(\theta)\right\vert ^{2}\propto\cos^{2}\left( \frac{1}{2}%
ka\theta\right)  \label{2slit}
\end{equation}
in the distribution of single-particle detections at angular deviation $%
\theta$ (measured from the normal to the screen). If the experiment is
performed with one particle at a time, each outcome $\theta$ will (in a
hidden-variables theory) be determined by a mapping $\theta=\theta(M,%
\lambda) $ (where again $M$ specifies the experimental arrangement). The
quantum distribution (\ref{2slit}) will correspond to the quantum
equilibrium distribution $\rho_{\mathrm{QT}}(\lambda)$, while nonequilibrium 
$\rho (\lambda)\neq\rho_{\mathrm{QT}}(\lambda)$ will generally imply
deviations from (\ref{2slit}) --- for example, an anomalous blurring of the
interference fringes.

\item Malus' law for two-state systems. For example, for single photons
incident on a polariser, quantum theory predicts a modulation%
\begin{equation}
p_{\mathrm{QT}}^{+}(\Theta )=\frac{1}{2}\left( 1+P\cos 2\Theta \right)
\label{Malus}
\end{equation}%
of the transmission probability, where $P$ is the (ensemble) polarisation of
the beam and $\Theta $ is the angle of the polariser. (For $P=1$, $p_{%
\mathrm{QT}}^{+}(\Theta )=\cos ^{2}\Theta $.) As shown elsewhere \cite{Sig},
Malus' law (\ref{Malus}) is equivalent to the additivity of expectation
values for non-commuting observables in a two-state system, and such
additivity generically breaks down in quantum nonequilibrium. Deviations
from (\ref{Malus}) then provide a convenient signature of nonequilibrium.

\item Gaussian vacuum fluctuations. Standard quantum field theory predicts
that seemingly empty space is the seat of field fluctuations corresponding
to a Gaussian random process, with a specified variance for each mode $%
\mathbf{k}$. Quantum nonequilibrium for vacuum fields will generically imply
a departure from Gaussianity and deviations from the predicted variance (or
width) for each $\mathbf{k}$.
\end{itemize}

The possible breakdown of Malus' law deserves special comment. Any two-state
quantum system has observables $\hat{\sigma}\equiv \mathbf{m}\cdot \mathbf{%
\hat{\sigma}}$ taking values $\sigma =\pm 1$, where $\mathbf{m}$ is a unit
vector in Bloch space and $\mathbf{\hat{\sigma}}$ is a Pauli spin operator.
Quantum theory predicts that, for an ensemble with density operator $\hat{%
\rho}$, the probability $p_{\mathrm{QT}}^{+}(\mathbf{m})$ for an outcome $%
\sigma =+1$ of a quantum measurement of $\hat{\sigma}$ is given by%
\begin{equation}
p_{\mathrm{QT}}^{+}(\mathbf{m})=\frac{1}{2}\left( 1+\left\langle \hat{\sigma}%
\right\rangle \right) =\frac{1}{2}\left( 1+\mathbf{m}\cdot \mathbf{P}\right)
\ ,  \label{p+QT}
\end{equation}%
where $\mathbf{P}=\langle \mathbf{\hat{\sigma}}\rangle =\mathrm{Tr}\left( 
\hat{\rho}\mathbf{\hat{\sigma}}\right) $ is the mean polarisation. (For
photons, an angle $\theta $ on the Bloch sphere corresponds to a physical
angle $\Theta =\theta /2$.) It is specifically the linearity in $\mathbf{m}$
of the quantum expectation value%
\begin{equation*}
E_{\mathrm{QT}}(\mathbf{m})\equiv \left\langle \mathbf{m}\cdot \mathbf{\hat{%
\sigma}}\right\rangle =\mathrm{Tr}\left( \hat{\rho}\mathbf{m}\cdot \mathbf{%
\hat{\sigma}}\right) =\mathbf{m}\cdot \mathbf{P}
\end{equation*}%
that is equivalent to expectation additivity for incompatible observables.
The proof is straightforward \cite{Sig}. For an arbitrary unit vector $%
\mathbf{m}=\sum_{i}c_{i}\mathbf{m}_{i}$, where $\left\{ \mathbf{m}_{1},\ 
\mathbf{m}_{2},\ \mathbf{m}_{3}\right\} $ is an orthonormal basis in Bloch
space, expectation additivity implies that $E_{\mathrm{QT}}(\mathbf{m}%
)=\sum_{i}c_{i}E_{\mathrm{QT}}(\mathbf{m}_{i})$. Invariance of $E_{\mathrm{QT%
}}(\mathbf{m})$ under a change of basis $\mathbf{m}_{i}\rightarrow \mathbf{m}%
_{i}^{\prime }$ then implies that $E_{\mathrm{QT}}(\mathbf{m})=\mathbf{m}%
\cdot \mathbf{P}$ where $\mathbf{P}\equiv \sum_{i}E_{\mathrm{QT}}(\mathbf{m}%
_{i})\mathbf{m}_{i}$ is a vector with norm $0\leq P\leq 1$. Using
expectation additivity again, we have $\mathbf{P}=\langle \mathbf{\hat{\sigma%
}}\rangle $.

A deterministic hidden-variables theory applied to a two-state system will
provide a mapping $\sigma =\sigma \left( \mathbf{m},\lambda \right) $ that
determines the measurement outcomes $\sigma =\pm 1$. As shown in ref. \cite%
{Sig}, for an arbitrary distribution $\rho (\lambda )\neq \rho _{\mathrm{QT}%
}(\lambda )$ of hidden variables $\lambda $ the nonequilibrium expectation
value%
\begin{equation*}
E(\mathbf{m})\equiv \left\langle \sigma \left( \mathbf{m},\lambda \right)
\right\rangle =\int d\lambda \ \rho (\lambda )\sigma \left( \mathbf{m}%
,\lambda \right)
\end{equation*}%
will generally \textit{not} take the linear form $\mathbf{m}\cdot \mathbf{P}$
for some Bloch vector $\mathbf{P}$, and the nonequilibrium outcome
probability%
\begin{equation}
p^{+}(\mathbf{m})=\frac{1}{2}\left( 1+E(\mathbf{m})\right)  \label{p+}
\end{equation}%
will generally not take the quantum form (\ref{p+QT}). Both the linearity
and the additivity are generically violated in quantum nonequilibrium.

A natural parameterisation of nonequilibrium outcome probabilities $p^{+}(%
\mathbf{m})$ for two-state systems may be obtained by expanding $p^{+}(%
\mathbf{m})$ in terms of spherical harmonics, with the unit vector $\mathbf{m%
}$ specified by angular coordinates $(\theta ,\phi )$ on the Bloch sphere.
For example, a probability law that includes a quadrupole term,%
\begin{equation}
p^{+}(\mathbf{m})=\frac{1}{2}\left( 1+\mathbf{m}\cdot \mathbf{P}+(\mathbf{m}%
\cdot \mathbf{b})(\mathbf{m}\cdot \mathbf{P})\right)  \label{QUAD}
\end{equation}%
(for some non-zero vector $\mathbf{b}$), corresponding to a \textit{nonlinear%
} expectation value%
\begin{equation}
E(\mathbf{m})=\mathbf{m}\cdot \mathbf{P}+(\mathbf{m}\cdot \mathbf{b})(%
\mathbf{m}\cdot \mathbf{P})\ ,
\end{equation}%
would signal a failure of expectation additivity and a violation of quantum
theory.

More generally, one might consider nonlinear expectation values%
\begin{equation*}
E(\mathbf{m})=m_{i}P_{i}+m_{i}m_{j}Q_{ij}+m_{i}m_{j}m_{k}R_{ijk}+\ ....
\end{equation*}%
(summing over repeated indices), where $Q_{ij}$, $R_{ijk}$, .... are tensors
in Bloch space. The experimental challenge is to set upper bounds on the
magnitudes $\left\vert Q_{ij}\right\vert $, $\left\vert R_{ijk}\right\vert $%
, .... , for systems in extreme conditions. The theoretical challenge, of
course, is to provide precise predictions for $Q_{ij}$, $R_{ijk}$, .... .

The statistical predictions of quantum theory and of quantum field theory
have of course been verified in countless experiments. For two-state
systems, for example, all known experimental data are consistent with $%
Q_{ij}=R_{ijk}=\ ....\ =0$. From a hidden-variables perspective, however,
there are good reasons to \textit{expect} that the experiments performed so
far yield agreement with quantum theory. This is because all the experiments
performed so far have been done with systems that have had a long and
violent astrophysical history. Atoms in the laboratory, for example, have a
history stretching back to the formation of stars, or even earlier (to big
bang nucleosynthesis), during which these atoms have undergone numerous
complex interactions with other systems. Every degree of freedom we have
access to has a complex past history of interaction with other degrees of
freedom, a history that ultimately merges with the history of the early
universe. This fact is highly significant, because it suggests that the
quantum equilibrium distribution $\rho _{\mathrm{QT}}(\lambda )$ observed
today could have emerged from past interactions, via a process of relaxation
(analogous to relaxation to thermal equilibrium in ordinary physics).

Relaxation to quantum equilibrium has been studied in some detail for the
case of pilot-wave theory. The quantity $H=\int dX\;P\ln(P/\left\vert
\Psi\right\vert ^{2})$ (equal to minus the relative entropy of an arbitrary
distribution $P$ with respect to $\left\vert \Psi\right\vert ^{2}$) obeys a
coarse-graining $H$-theorem analogous to the classical one \cite%
{AV91a,AV92,AV01}; and numerical simulations for simple two-dimensional
systems \cite{VW05,Sky} show a rapid (approximately exponential) decay of
the coarse-grained $H$-function, $\bar{H}(t)\rightarrow0$, with a
corresponding coarse-grained relaxation $\bar{P}\rightarrow\overline{%
\left\vert \Psi\right\vert ^{2}}$ (given appropriate initial conditions on $%
P $ and $\Psi $, see ref. \cite{AV01}).

In pilot-wave theory, then, given the known past history of the universe,
there is every reason to expect the systems being examined today to be in
quantum equilibrium. Presumably, similar conclusions would hold in any
reasonable (deterministic) hidden-variables theory: we expect that the known
past interactions will generate a similar relaxation $\rho(\lambda
)\rightarrow\rho_{\mathrm{QT}}(\lambda)$.

In this scenario, quantum theory is merely an effective theory, describing
the physics of an equilibrium state that emerged some time in the remote
past. Considering this scenario further suggests clues as to where quantum
theory might break down.

The obvious place to look is the very early universe. At sufficiently early
times, quantum nonequilibrium $\rho(\lambda)\neq\rho_{\mathrm{QT}}(\lambda)$
may have still existed. How can one probe such early times experimentally?
One possibility is provided by inflationary cosmology, according to which
primordial vacuum fluctuations in a scalar field $\phi$ (present during an
early period of exponential spatial expansion) are responsible for the early
inhomogeneities that seeded the formation of large-scale structure in the
universe and that left an observable imprint on the cosmic microwave
background (CMB). This suggests that primordial quantum nonequilibrium could
have a measurable effect on the CMB temperature anisotropy \cite{AV06}.
Another possibility is based on the idea \cite{AV01} that certain particle
species may have decoupled so early that they did not have time to reach
quantum equilibrium: such nonequilibrium relic particles could still exist
today. One is then led to consider testing quantum theory for relic
particles from very early times.

Instead of looking for residual nonequilibrium from the distant past, would
it be possible to \textit{generate} nonequilibrium today? It has been
suggested \cite{AV04} that gravitation may be capable of generating quantum
nonequilibrium. In particular, information loss in black holes might be
avoided if Hawking radiation consisted of nonequilibrium particles, since
the final state could then contain more information than the conventional
(quantum) final state. Following this line of reasoning, one is led to
suggest that if one half of a bipartite entangled state fell behind the
event horizon of a black hole, the other half would evolve away from quantum
equilibrium. Such a situation might occur naturally via atomic cascade
emissions in black-hole accretion discs.

There are also theoretical reasons for suspecting that quantum-gravitational
effects could induce pure-to-mixed transitions in, for example, oscillating
neutrinos. Motivated once again by the possible avoidance of information
loss, such transitions might be accompanied by the generation of quantum
nonequilibrium.

Finally, the possibility of gravitational effects generating nonequilibrium
at the Planck scale motivates us to consider tests of quantum probabilities
at very small lengthscales, for all particles whatever their origin. As we
shall see, in the right circumstances the spreading of wave packets for
particles emitted by remote sources can act as a cosmological `microscope',
expanding tiny deviations from quantum theory to observable scales.

We shall now examine these suggestions in turn.

\section{Inflation as a Test of Quantum Theory in the Early Universe}

The temperature anisotropy $\Delta T(\theta,\phi)\equiv T(\theta,\phi)-T$ of
the microwave sky may be expanded in terms of spherical harmonics as%
\begin{equation}
\frac{\Delta T(\theta,\phi)}{T}=\sum_{l=2}^{\infty}%
\sum_{m=-l}^{+l}a_{lm}Y_{lm}(\theta,\phi)\ .
\end{equation}
It is usual to regard the observed $T(\theta,\phi)$ as a realisation of a
stochastic process, such that the underlying probability distribution for
each coefficient $a_{lm}$ is independent of $m$ (as follows if the
probability distribution for $T(\theta,\phi)$ is assumed to be rotationally
invariant). For large enough $l$, the (theoretical) ensemble average $%
\left\langle \left\vert a_{lm}\right\vert ^{2}\right\rangle $ may then be
accurately estimated as%
\begin{equation}
\left\langle \left\vert a_{lm}\right\vert ^{2}\right\rangle \approx\frac {1}{%
2l+1}\sum_{m=-l}^{+l}\left\vert a_{lm}\right\vert ^{2}\equiv C_{l}\ .
\end{equation}
The anisotropy $\Delta T(\theta,\phi)$ is believed to have been produced by
(classical) inhomogeneities on the last scattering surface (when CMB photons
decoupled from matter). There is a well-established theory expressing the $%
a_{lm}$ in terms of a Fourier-transformed `primordial curvature
perturbation' $\mathcal{R}_{\mathbf{k}}$ (see, for example, refs. \cite%
{AV06,LL} for details). Assuming that the underlying probability
distribution for $\mathcal{R}_{\mathbf{k}}$ is translationally invariant, it
may be shown that%
\begin{equation}
C_{l}=\frac{1}{2\pi^{2}}\int_{0}^{\infty}\frac{dk}{k}\ \mathcal{T}^{2}(k,l)%
\mathcal{P}_{\mathcal{R}}(k)\ ,
\end{equation}
where $\mathcal{T}$ is a function encoding the relevant astrophysical
processes and%
\begin{equation}
\mathcal{P}_{\mathcal{R}}(k)\equiv\frac{4\pi k^{3}}{V}\left\langle
\left\vert \mathcal{R}_{\mathbf{k}}\right\vert ^{2}\right\rangle
\end{equation}
is the primordial power spectrum for $\mathcal{R}_{\mathbf{k}}$ (with $V$ a
normalisation volume). Data for the $C_{l}$ suggest that $\mathcal{P}_{%
\mathcal{R}}(k)\approx\mathrm{const.}$ (an approximately scale-free
spectrum) \cite{CMBdata}.

Now, inflationary cosmology predicts that $\mathcal{R}_{\mathbf{k}}$ is
given by \cite{LL}%
\begin{equation}
\mathcal{R}_{\mathbf{k}}=-\left[ \frac{H}{\dot{\phi}_{0}}\phi_{\mathbf{k}}%
\right] _{t=t_{\ast}(k)}\ ,
\end{equation}
where $H\equiv\dot{a}/a$ is the (approximately constant) Hubble parameter of
the inflating universe (with metric $d\tau^{2}=dt^{2}-a^{2}d\mathbf{x}^{2}$
and scale factor $a=a(t)$), $\phi_{0}$ and $\phi$ are respectively the
spatially homogeneous and inhomogeneous parts of the inflaton field, and the
right hand side is evaluated at a time $t_{\ast}(k)$ a few $e$-folds after
the (exponentially expanding) physical wavelength $\lambda_{\mathrm{phys}%
}=2\pi a(t)/k$ of the mode $\mathbf{k}$ exceeds (or `exits') the Hubble
radius $H^{-1}$. To a first approximation, inflation predicts that $\phi_{%
\mathbf{k}}$ will have (at time $t_{\ast}(k)$) a quantum variance%
\begin{equation}
\left\langle |\phi_{\mathbf{k}}|^{2}\right\rangle _{\mathrm{QT}}=\frac {V}{%
2(2\pi)^{3}}\frac{H^{2}}{k^{3}}  \label{BD}
\end{equation}
and a scale-invariant power spectrum%
\begin{equation}
\mathcal{P}_{\phi}^{\mathrm{QT}}(k)\equiv\frac{4\pi k^{3}}{V}\left\langle
\left\vert \phi_{\mathbf{k}}\right\vert ^{2}\right\rangle _{\mathrm{QT}}=%
\frac{H^{2}}{4\pi^{2}}  \label{HZ}
\end{equation}
(where $\left\langle \left\vert \phi_{\mathbf{k}}\right\vert
^{2}\right\rangle _{\mathrm{QT}}$ is obtained from the Bunch-Davies vacuum
in de Sitter space, for $\lambda_{\mathrm{phys}}>>H^{-1}$). This results in
a scale-free spectrum (in the slow-roll limit $\dot{H}\rightarrow0$) for $%
\mathcal{R}_{\mathbf{k}}$,%
\begin{equation}
\mathcal{P}_{\mathcal{R}}^{\mathrm{QT}}(k)=\frac{1}{4\pi^{2}}\left[ \frac{%
H^{4}}{\dot{\phi}_{0}^{2}}\right] _{t_{\ast}(k)}\ ,  \label{PRQT}
\end{equation}
in approximate agreement with what is observed.

Quantum nonequilibrium in the Bunch-Davies vacuum would yield deviations
from (\ref{BD}). Further, in the pilot-wave version of quantum field theory,
it may be shown \cite{AV06} that any (microscopic) quantum nonequilibrium
that is present at the onset of inflation will be \textit{preserved} during
the inflationary phase (instead of relaxing), and will in fact be
transferred to macroscopic lengthscales by the growth of physical
wavelengths $\lambda _{\mathrm{phys}}\propto a(t)\propto e^{Ht}$.

This is shown by calculating the de Broglie-Bohm trajectories for the
inflaton field. Writing $\phi _{\mathbf{k}}=\frac{\sqrt{V}}{(2\pi )^{3/2}}%
\left( q_{\mathbf{k}1}+iq_{\mathbf{k}2}\right) $ (for real $q_{\mathbf{k}r}$%
, $r=1$, $2$), the Bunch-Davies wave functional takes the product form $\Psi
\lbrack q_{\mathbf{k}r},t]=\prod\limits_{\mathbf{k}r}\psi _{\mathbf{k}r}(q_{%
\mathbf{k}r},t)$, and the de Broglie equation of motion for $q_{\mathbf{k}r}$
is%
\begin{equation*}
\frac{dq_{\mathbf{k}r}}{dt}=\frac{1}{a^{3}}\frac{\partial s_{\mathbf{k}r}}{%
\partial q_{\mathbf{k}r}}\ ,
\end{equation*}%
where $\psi _{\mathbf{k}r}=\left\vert \psi _{\mathbf{k}r}\right\vert e^{is_{%
\mathbf{k}r}}$. Using the known form for $\psi _{\mathbf{k}r}$, it is found
that%
\begin{equation*}
\frac{dq_{\mathbf{k}r}}{dt}=-\frac{k^{2}Hq_{\mathbf{k}r}}{k^{2}+H^{2}a^{2}}\
,
\end{equation*}%
which has the solution%
\begin{equation*}
q_{\mathbf{k}r}(\eta )=q_{\mathbf{k}r}(0)\sqrt{1+k^{2}\eta ^{2}}\ ,
\end{equation*}%
where $\eta =-1/Ha$ is the conformal time (running from $-\infty $ to $0$).
Given this solution for the trajectories, one may easily construct the exact
evolution of an arbitrary distribution $\rho _{\mathbf{k}r}(q_{\mathbf{k}%
r},\eta )$ (generally $\neq \left\vert \psi _{\mathbf{k}r}(q_{\mathbf{k}%
r},\eta )\right\vert ^{2}$). The time evolution amounts to a homogeneous
contraction of both $\left\vert \psi _{\mathbf{k}r}\right\vert ^{2}$ and $%
\rho _{\mathbf{k}r}$. At times $\eta <0$, $\left\vert \psi _{\mathbf{k}%
r}\right\vert ^{2}$ is a contracting Gaussian packet of width $\Delta _{%
\mathbf{k}r}(\eta )=\Delta _{\mathbf{k}r}(0)\sqrt{1+k^{2}\eta ^{2}}$. In the
late-time limit $\eta \rightarrow 0$, $\left\vert \psi _{\mathbf{k}%
r}\right\vert ^{2}$ approaches a static Gaussian of width $\Delta _{\mathbf{k%
}r}(0)=H/\sqrt{2k^{3}}$. At times $\eta <0$, $\rho _{\mathbf{k}r}$ is a
contracting arbitrary distribution of width $D_{\mathbf{k}r}(\eta )=D_{%
\mathbf{k}r}(0)\sqrt{1+k^{2}\eta ^{2}}$ (with arbitrary $D_{\mathbf{k}r}(0)$%
). In the late-time limit $\eta \rightarrow 0$, $\rho _{\mathbf{k}r}$
approaches a static packet of width $D_{\mathbf{k}r}(0)$ (the asymptotic
packet differing from the earlier packet by a homogeneous rescaling of $q$).
We then have the result%
\begin{equation}
\frac{D_{\mathbf{k}r}(t)}{\Delta _{\mathbf{k}r}(t)}=(\mathrm{const.\ in\ time%
})\equiv \sqrt{\xi (k)}\ ,
\end{equation}%
where for simplicity we assume that (like $\Delta _{\mathbf{k}r}$) the
nonequilibrium width $D_{\mathbf{k}r}$ depends on $k$ and $t$ only. (For
each mode, the factor $\xi (k)$ may be defined at any convenient fiducial
time.) Thus, for each mode $\mathbf{k}$, the widths of the nonequilibrium
and equilibrium distributions remain in a fixed ratio over time.

Thus, at least to a first approximation (treating the inflationary phase as
an exact de Sitter expansion), if quantum nonequilibrium exists at early
times it will not relax during the inflationary phase. Instead, it will
indeed be preserved, and be transferred to macroscopic scales by the
expansion of physical wavelengths $\lambda _{\mathrm{phys}}$. This process
is especially striking in the late-time limit, where both $\rho _{\mathbf{k}%
r}$ and $\left\vert \psi _{\mathbf{k}r}\right\vert ^{2}$ become static. Once
the mode exits the Hubble radius, the nonequilibrium becomes `frozen', while 
$\lambda _{\mathrm{phys}}$ continues to grow exponentially. The `frozen'
nonequilibrium then corresponds to a physical lengthscale that grows
exponentially with time, from microscopic to macroscopic scales. And of
course, once inflation has ended, curvature perturbations $\mathcal{R}_{%
\mathbf{k}}$ at macroscopic lengthscales are transferred to cosmological
lengthscales by the subsequent (post-inflationary) Friedmann expansion.

Writing the nonequilibrium variance as%
\begin{equation}
\left\langle |\phi_{\mathbf{k}}|^{2}\right\rangle =\left\langle |\phi _{%
\mathbf{k}}|^{2}\right\rangle _{\mathrm{QT}}\xi(k)\ ,
\end{equation}
the resulting power spectrum for $\mathcal{R}_{\mathbf{k}}$ is then just the
usual result (\ref{PRQT}) multiplied by the factor $\xi(k)$:%
\begin{equation}
\mathcal{P}_{\mathcal{R}}(k)=\frac{\xi(k)}{4\pi^{2}}\left[ \frac{H^{4}}{\dot{%
\phi}_{0}^{2}}\right] _{t_{\ast}(k)}\ .  \label{PR}
\end{equation}
Early quantum nonequilibrium will generally break the scale invariance of
the primordial power spectrum $\mathcal{P}_{\mathcal{R}}(k)$ (at least in
pilot-wave theory). Measurements of the angular power spectrum $C_{l}$ of
the microwave sky may be used --- in the context of inflationary theory ---
to constrain the primordial `nonequilibrium function' $\xi(k)$ \cite{AV06}.

Other measurable effects of early nonequilibrium include violation of the
scalar-tensor consistency relation, non-Gaussianity, and non-random
primordial phases \cite{AV06}.

Work in progress attempts to predict features of the function $\xi(k)$, by
studying the evolution of nonequilibrium in an assumed pre-inflationary era:
preliminary results suggest that, at the beginning of inflation,
nonequilibrium is more likely to have survived at large wavelengths (small $%
k $).

Note that primordial nonequilibrium $\xi (k)\neq 1$ might be generated
during the inflationary phase by novel gravitational effects at the Planck
scale (see section 5) --- as well as, or instead of, being a remnant of an
earlier nonequilibrium epoch.

Finally, we remark that Perez \textit{et al}. \cite{PSS06} have also
considered modifying quantum theory in an inflationary context. Their
primary motivation is the quantum measurement problem (which is of course
especially severe in cosmology). In particular, they discuss how predictions
for the CMB could be affected by a dynamical collapse of the wave function
in the early universe.

\section{Relic Nonequilibrium Particles}

The early universe contains a mixture of effectively massless (relativistic)
particles. According to the standard analysis, relaxation to thermal
equilibrium between different particle species depends on two competing
effects: interactions driving different species towards mutual equilibrium,
and spatial expansion making different species fall out of mutual
equilibrium. Relaxation occurs only if the former overcomes the latter, that
is, only if the mean free time $t_{\func{col}}$ between collisions is
smaller than the timescale $t_{\exp }\equiv a/\dot{a}$ of spatial expansion.
In a Friedmann model (perhaps pre- or post-inflationary), $a\propto t^{1/2}$
and $t_{\exp }\propto 1/T^{2}$ (where $T$ is the photon temperature). Thus,
if $t_{\func{col}}=t_{\func{col}}(T)$ falls off slower than $1/T^{2}$, at
sufficiently high temperatures $t_{\func{col}}\gtrsim t_{\exp }$ and thermal
equilibrium between the species will not be achieved --- or at least, not
until the temperature has dropped sufficiently for $t_{\func{col}}\lesssim
t_{\exp }$ to hold. Similarly, species that are in thermal equilibrium will
subsequently decouple if $t_{\func{col}}$ becomes larger than $t_{\exp }$ as
the universe expands and $T$ decreases (as occurs for CMB photons at
recombination). The thermal history of the universe then depends crucially
on the functions $t_{\func{col}}(T)$, which in turn depend on the relevant
scattering cross sections.

We expect that relaxation to quantum equilibrium in an expanding universe
will likewise depend on two competing effects: the usual relaxation seen in
flat spacetime, and the stretching of the nonequilibrium lengthscale caused
by spatial expansion \cite{AV01}.

As already mentioned, numerical simulations in pilot-wave theory show a very
efficient relaxation for systems with two degrees of freedom (given
appropriate initial conditions). These simulations were carried out on a
static background (flat) spacetime, with a wave function equal to a
superposition of many different energy eigenstates, for nonrelativistic
particles in a two-dimensional box \cite{VW05} and in a two-dimensional
harmonic oscillator potential \cite{Sky}. The latter case is mathematically
equivalent to that of a single decoupled mode $\mathbf{k}$ of a free scalar
field on Minkowski spacetime: again writing $\phi _{\mathbf{k}}=\frac{\sqrt{V%
}}{(2\pi )^{3/2}}\left( q_{\mathbf{k}1}+iq_{\mathbf{k}2}\right) $ as above,
the wave function $\psi _{\mathbf{k}}=\psi _{\mathbf{k}}(q_{\mathbf{k}1},q_{%
\mathbf{k}2},t)$ of the mode satisfies%
\begin{equation}
i\frac{\partial \psi _{\mathbf{k}}}{\partial t}=-\frac{1}{2}\left( \frac{%
\partial ^{2}}{\partial q_{\mathbf{k}1}^{2}}+\frac{\partial ^{2}}{\partial
q_{\mathbf{k}2}^{2}}\right) \psi _{\mathbf{k}}+\frac{1}{2}k^{2}\left( q_{%
\mathbf{k}1}^{2}+q_{\mathbf{k}2}^{2}\right) \psi _{\mathbf{k}}\ ,
\end{equation}%
and the de Broglie velocities for $q_{\mathbf{k}r}$ are $\dot{q}_{\mathbf{k}%
r}=\partial s_{\mathbf{k}}/\partial q_{\mathbf{k}r}$ (with $\psi _{\mathbf{k}%
}=\left\vert \psi _{\mathbf{k}}\right\vert e^{is_{\mathbf{k}}}$), just as in
the pilot-wave theory of a nonrelativistic particle of unit mass in a
harmonic oscillator potential in the $q_{\mathbf{k}1}-q_{\mathbf{k}2}$
plane. Thus we deduce that, in the absence of gravity, for a single mode $%
\mathbf{k}$ in a superposition of many different states of definite
occupation number, the probability distribution $\rho _{\mathbf{k}}(q_{%
\mathbf{k}1},q_{\mathbf{k}2},t)$ will rapidly relax to equilibrium, $\rho _{%
\mathbf{k}}\rightarrow \left\vert \psi _{\mathbf{k}}\right\vert ^{2}$ (on a
coarse-grained level, again given appropriate initial conditions).

Now, in a flat expanding universe, again with metric $d\tau
^{2}=dt^{2}-a^{2}d\mathbf{x}^{2}$, the pilot-wave equations for a decoupled
mode become \cite{AV06}%
\begin{equation}
i\frac{\partial \psi _{\mathbf{k}}}{\partial t}=-\frac{1}{2a^{3}}\left( 
\frac{\partial ^{2}}{\partial q_{\mathbf{k}1}^{2}}+\frac{\partial ^{2}}{%
\partial q_{\mathbf{k}2}^{2}}\right) \psi _{\mathbf{k}}+\frac{1}{2}%
ak^{2}\left( q_{\mathbf{k}1}^{2}+q_{\mathbf{k}2}^{2}\right) \psi _{\mathbf{k}%
}  \label{An}
\end{equation}%
and%
\begin{equation}
\dot{q}_{\mathbf{k}1}=\frac{1}{a^{3}}\frac{\partial s_{\mathbf{k}}}{\partial
q_{\mathbf{k}1}},\ \ \ \ \dot{q}_{\mathbf{k}2}=\frac{1}{a^{3}}\frac{\partial
s_{\mathbf{k}}}{\partial q_{\mathbf{k}2}}\ .  \label{B}
\end{equation}%
How does the presence of $a=a(t)$ affect the time evolution? If $\lambda _{%
\mathrm{phys}}<<H^{-1}$, we recover the Minkowski-space evolution --- the
expansion timescale $H^{-1}\equiv a/\dot{a}$ being much larger than the
timescale $\sim \lambda _{\mathrm{phys}}$ (with $c=1$) over which $\psi _{%
\mathbf{k}}$ evolves --- and so a superposition of many different states of
definite occupation number (for the mode $\mathbf{k}$) will again rapidly
relax to equilibrium. On the other hand, if $\lambda _{\mathrm{phys}%
}>>H^{-1} $, we expect $\psi _{\mathbf{k}}$ and the associated de
Broglie-Bohm trajectories to be approximately static over timescales such
that $\lambda _{\mathrm{phys}}\propto a(t)$ expands significantly, so that
relaxation is suppressed. The spatial expansion then results in a transfer
of nonequilibrium to larger lengthscales (as we saw in late-time inflation).

There are then two `competing' effects: the usual relaxation to equilibrium,
and the transfer of nonequilibrium to larger lengthscales. The former
dominates for $\lambda _{\mathrm{phys}}<<H^{-1}$, the latter for $\lambda _{%
\mathrm{phys}}>>H^{-1}$. In a radiation-dominated phase, with $a\propto
t^{1/2}$, we have $\lambda _{\mathrm{phys}}\propto t^{1/2}$ and $%
H^{-1}\propto t$. Thus, at sufficiently small times, \textit{all} physical
wavelengths are larger than the Hubble radius ($\lambda _{\mathrm{phys}%
}>H^{-1}$), and the above reasoning suggests that relaxation to equilibrium
will be suppressed (until later times when $\lambda _{\mathrm{phys}}$
becomes smaller than $H^{-1}$). While further study is needed --- such as
numerical simulations based on (\ref{An}), (\ref{B}), and consideration of
entangled and also mixed states --- we seem to have a mechanism whereby
spatial expansion at very early times can suppress the normal relaxation to
equilibrium.

Similar conclusions have been arrived at in terms of the pilot-wave theory
of particles \cite{AV01}. If the distribution of particle positions contains
nonequilibrium below a certain lengthscale, the spatial expansion will
transfer the nonequilibrium to larger lengthscales. Further, a simple
estimate $\tau\sim\hslash/kT$ of the relaxation timescale suggests that
relaxation will be suppressed when $\tau\gtrsim t_{\exp}\sim(1\sec)(1\ 
\mathrm{MeV}/kT)^{2}$ --- that is, when $kT\gtrsim10^{18}\ \mathrm{GeV}%
\approx0.1kT_{\mathrm{P}}$ or $t\lesssim10t_{\mathrm{P}}$ (where $T_{\mathrm{%
P}}$ and $t_{\mathrm{P}}$ are respectively the Planck temperature and time).
We emphasise that this estimate, while suggestive, is only heuristic.

If relaxation to quantum equilibrium is indeed suppressed at sufficiently
early times, in a realistic cosmological model, this raises the exciting
possibility that if the universe indeed began in a state of quantum
nonequilibrium, then remnants of such nonequilibrium could have survived to
the present day --- for particles that decoupled at times so early that
equilibrium had not yet been reached. Relic gravitons are believed to
decouple at $T\sim T_{\mathrm{P}}$, and there may well be other, more exotic
particles (associated with physics beyond the standard model) that decoupled
soon after $T_{\mathrm{P}}$. A subsequent inflationary era would presumably
dilute their density beyond any hope of detection, but in the absence of
inflation it is possible that such particles could have a significant
abundance today. Further, such relic nonequilibrium particles might
annihilate or decay, producing nonequilibrium photons --- which could be
detected directly, and tested for violations of Malus' law or for anomalous
diffraction and interference patterns.

\section{Tests of Quantum Theory with Black Holes}

According to pilot-wave theory, once quantum equilibrium is reached it is
not possible to escape from it (leaving aside the remote possibility of rare
fluctuations \cite{AV92}). A universe in quantum equilibrium is then
analogous to a universe stuck in a state of global thermal equilibrium or
thermodynamic `heat death'. Further, in quantum equilibrium it is not
possible to harness nonlocality for signalling, just as in global thermal
equilibrium it is not possible to convert heat into work [5--8].

However, pilot-wave theory has been well developed only for
non-gravitational physics. Indeed, despite much effort, standard quantum
theory too has yet to be extended to gravity. It is then conceivable that
quantum equilibrium as we know it will turn out to be gravitationally
unstable: in a future hidden-variables theory incorporating gravitation,
there could exist processes that \textit{generate} quantum nonequilibrium.

One such process might be the formation and evaporation of a black hole,
which arguably allows a pure quantum state to evolve into a mixed one \cite%
{H76}. It has been suggested that the resulting `information loss' (the
inability in principle to retrodict the initial state from the final one)
could be avoided if the outgoing Hawking radiation were in a state of
quantum nonequilibrium, enabling it to carry more information than
conventional radiation could \cite{AV04}. A mechanism has been suggested,
whereby (putative) nonequilibrium behind the event horizon is transmitted to
the exterior region via the entanglement between the ingoing and outgoing
radiation modes \cite{AV04}. It has also been proposed that the decreased
`hidden-variable entropy' $S_{\mathrm{hv}}$ (minus the subquantum $H$%
-function, suitably generalised to mixed states \cite{AV04}) of the outgoing
nonequilibrium radiation should balance the increase in von Neumann entropy $%
S_{\mathrm{vonN}}=-\mathrm{Tr}(\hat{\rho}\ln \hat{\rho})$ associated with
the pure-to-mixed transition:%
\begin{equation}
\Delta \left( S_{\mathrm{hv}}+S_{\mathrm{vonN}}\right) =0\ .  \label{ConsS}
\end{equation}

At the time of writing, the proposed conservation rule (\ref{ConsS}) is only
a simple and somewhat arbitrary hypothesis, relating as it does two very
different kinds of entropy, $S_{\mathrm{hv}}$ and $S_{\mathrm{vonN}}$
(though it has been shown \cite{AV04} that these entropies must be related
even in non-gravitational processes, in ways that need to be explored
further). If the pure-to-mixed transition does indeed generate
nonequilibrium, it might be hoped that (\ref{ConsS}) will hold at least as
an order-of-magnitude estimate.

The above (obviously speculative) idea could be tested, should Hawking
radiation from microscopic black holes ever be observed. Primordial black
holes of mass $M\sim 10^{15}\ \mathrm{g}$ are expected to be evaporating
today, producing (among other particles) gamma-rays peaked at $\sim 100$ $%
\mathrm{MeV}$ \cite{Page}. Such radiation has been searched for, so far with
no definitive detection, and further searches are under way. Should $\gamma $%
-rays from the evaporation of primordial black holes ever be detected, we
propose that their polarisation probabilities be carefully checked (for
example by Compton polarimetry) for deviations from the standard modulation (%
\ref{Malus}). Another possibility, according to theories with large extra
dimensions \cite{AHDD}, is that microscopic black holes could be produced in
collisions at the $\mathrm{TeV}$ scale. If so, their decay products could be
tested for deviations from (\ref{Malus}).

If the entanglement between ingoing and outgoing Hawking radiation modes
does indeed provide a channel for nonlocal information flow from behind the
horizon, then one would expect a similar process to occur if, for an
`EPR-pair' initially in the exterior region, one half of the entangled state
fell behind the horizon. For an ensemble of such pairs, the particles left
in the exterior region should evolve away from quantum equilibrium --- by an
amount that can be estimated from the proposed rule (\ref{ConsS}) (where $%
\Delta S_{\mathrm{vonN}}$ is obtained by tracing over the infalling
particles).

It has been argued that, if the information loss envisaged by Hawking is to
be avoided by some form of nonlocal information flow, then such flow must
occur even while the hole is still macroscopic \cite{Dan93}. Similar
arguments lead us to conclude that, even for a \textit{macroscopic} black
hole, allowing one half of an EPR-pair to fall behind the horizon will cause
the other half to evolve away from quantum equilibrium --- over a timescale
small compared to the evaporation timescale \cite{AV04}.

This motivates us to propose another test. Most galactic nuclei contain a
supermassive black hole ($M\sim 10^{6}-10^{10}\ M_{\odot }$) surrounded by a
thin accretion disc \cite{FF05}. It is well-established that X-ray emission
lines, in particular the K$\alpha $ iron line at $6.4$ $\mathrm{keV}$, may
be used to probe the spacetime geometry in the strong gravity region close
to the event horizon \cite{RN03}. The intrinsically narrow line is broadened
and skewed by relativistic effects, with an extended red wing caused by the
gravitational redshift of photons emitted from very near the horizon. This
much is well known. Now, the idea is to identify an atomic \textit{cascade}
emission that generates entangled photon pairs at small radii, such that a
significant fraction of the photons reaching Earth have partners that fell
behind the horizon. Polarisation measurements of the received photons would
then provide a test of Malus' law (\ref{Malus}) --- and a probe of possible
nonequilibrium, for example in the form of a quadrupole probability law (\ref%
{QUAD}) --- for photons entangled with partners inside the black hole.

The feasibility of this experiment has been discussed in detail elsewhere 
\cite{AV04,AV04b}. Here, we summarise what appear to be the main points:

\begin{itemize}
\item In a $0-1-0$ two-photon cascade, for example, the polarisation state
shows a strong and phase-coherent entanglement only if the emitted momenta
are approximately antiparallel \cite{Asp02}. This may be realised in our
experiment by restricting attention to photons with the largest redshift:
these have emission radii $r_{\mathrm{e}}$ closest to the horizon at $%
r_{+}=M+\sqrt{M^{2}-a^{2}}$ (where $a$ is the specific angular momentum of
the hole), and as $r_{\mathrm{e}}\rightarrow r_{+}$ the photons will escape
--- and avoid being absorbed by the hole or the accretion disc --- only if
they are directed parallel to the surface of the disc \cite{C75}.

\item The effect will be \textit{diluted} by received photons with: (a) no
cascade partners, (b) cascade partners that were not captured by the black
hole, (c) cascade partners that were captured but did not have appropriately
directed momenta at the point of emission.

\item Scattering along the line of sight could degrade the entanglement
between the outgoing and ingoing photons, and might cause relaxation $%
\rho(\lambda)\rightarrow\rho_{\mathrm{QT}}(\lambda)$. This may be minimised
by an appropriate choice of photon frequency and by choosing an accretion
disc viewed face-on (with a clear line of sight to the central black hole).

\item If the nonequilibrium distribution $\rho(\lambda)\neq\rho_{\mathrm{QT}%
}(\lambda)$ for the received photons depends on the spatial location of the
emission, the sought-for effect could be smeared out by spatial averaging
over the emitting region. If instead $\rho(\lambda)$ is independent of
location, such averaging will have no effect.

\item Only about $0.6\%$ of the observed K$\alpha$ photons are expected to
have L$\alpha$ cascade partners \cite{AV04,AV04b}. We hope that other
relativistically broadened lines will be discovered, with a larger fraction
of cascade partners.\footnote{%
Broadened lines from oxygen, nitrogen and carbon have in fact already been
reported \cite{Ogle04}.}
\end{itemize}

Note that true deviations from (\ref{Malus}) may be distinguished from
ordinary noise and experimental errors by comparing results from the
astronomical source with results from a laboratory source. Also, if the
effect exists, it will be larger towards the red end of the (broadened)
emission line, because these photons are emitted closer to the horizon and
are therefore more likely to have partners that were captured.

\section{Neutrino Oscillations}

Microscopic quantum-gravitational effects might induce a pure-to-mixed
evolution of the quantum state in a system of oscillating neutrinos,
resulting in damping and decoherence effects that might be observable over
astrophysical and cosmological (or even just atmospheric) path lengths ---
see, for example, refs. [33--39]. Such evolution may be modelled by
corrections to the usual unitary evolution of the density operator $\hat{\rho%
}(t)$. Writing%
\begin{equation*}
\frac{d\hat{\rho}}{dt}=-i[\hat{H},\hat{\rho}]-\mathcal{D}(\hat{\rho})\ ,
\end{equation*}%
the extra term $\mathcal{D}$ breaks the usual conservation of $\mathrm{Tr}(%
\hat{\rho}^{2})$. It is usually assumed that $\mathcal{D}$ takes a Lindblad
form, and that the mean energy $\mathrm{Tr}(\hat{\rho}\hat{H})$ is
conserved. Under the usual assumptions, the term $\mathcal{D}$ generates an
increase in von Neumann entropy $S_{\mathrm{vonN}}=-\mathrm{Tr}(\hat{\rho}%
\ln \hat{\rho})$ over time. (See, for example, ref. \cite{Lisi00}.)

A detailed phenomenological parameterisation of $\mathcal{D}$ has been
developed, and extensive comparisons with data have been made \cite%
{Lisi00,Mav04,Mor06,Bar06}. If we follow the hypotheses of section 5
(assuming that $\mathcal{D}$ originates, for example, from the formation and
evaporation of microscopic black holes), then any such pure-to-mixed
transition will generate quantum nonequilibrium, of a magnitude that might
be constrained by (\ref{ConsS}). This will result in nonequilibrium
anomalies in the composition of an oscillating neutrino beam.

Consider the simple case of just two flavours, labelled $\nu _{\mu }$ and $%
\nu _{\tau }$. Lepton number eigenstates $\left\vert \nu _{\mu
}\right\rangle $, $\left\vert \nu _{\tau }\right\rangle $ are linear
combinations%
\begin{align*}
\left\vert \nu _{\mu }\right\rangle & =\left\vert \nu _{1}\right\rangle \cos
\alpha +\left\vert \nu _{2}\right\rangle \sin \alpha \ , \\
\left\vert \nu _{\tau }\right\rangle & =-\left\vert \nu _{1}\right\rangle
\sin \alpha +\left\vert \nu _{2}\right\rangle \cos \alpha
\end{align*}%
of mass eigenstates (masses $m_{1}$, $m_{2}$) where $\alpha $ is the mixing
angle. For a beam of energy $E>>m_{1}$, $m_{2}$, terms in $\left\vert \nu
_{1}\right\rangle $ and $\left\vert \nu _{2}\right\rangle $ propagate with
relative phases $e^{ikt/2}$ and $e^{-ikt/2}$ respectively, where $k\equiv
(m_{2}^{2}-m_{1}^{2})/2E$ \cite{Beu03}.

The oscillating two-state system may be represented in Bloch space, with $%
\left\vert \nu _{1}\right\rangle $ and $\left\vert \nu _{2}\right\rangle $
corresponding to unit vectors respectively up and down the $z$-axis. We then
have a Hamiltonian $\hat{H}=-(k/2)\hat{\sigma}_{z}$ (where $\hat{\sigma}_{z}$
is a Pauli operator). For an arbitrary density operator $\hat{\rho}=\frac{1}{%
2}(\hat{I}+\mathbf{P}\cdot \mathbf{\hat{\sigma}})$, the mean polarisation $%
\mathbf{P}=\mathrm{Tr}(\hat{\rho}\mathbf{\hat{\sigma}})$ then evolves as $d%
\mathbf{P}/dt=\mathbf{k}\times \mathbf{P}$ where $\mathbf{k}\equiv (0,0,-k)$%
. An initial pure state $\hat{\rho}(0)=\left\vert \nu _{\mu }\right\rangle
\left\langle \nu _{\mu }\right\vert $ with%
\begin{equation}
\mathbf{P}(0)=(\sin 2\alpha ,0,\cos 2\alpha )  \label{P0}
\end{equation}%
evolves into a pure state with%
\begin{equation}
\mathbf{P}(t)=(\sin 2\alpha \cos kt,-\sin 2\alpha \sin kt,\cos 2\alpha )
\label{P1}
\end{equation}%
(where $\left\vert \mathbf{P}(t)\right\vert =1$), and the quantum survival
probability for $\nu _{\mu }$ shows the well-known oscillation%
\begin{equation*}
p_{\mathrm{QT}}^{\mu }(t)=\mathrm{Tr}(\hat{\rho}(t)\left\vert \nu _{\mu
}\right\rangle \left\langle \nu _{\mu }\right\vert )=1-\frac{1}{2}(1-\cos
kt)\sin ^{2}2\alpha
\end{equation*}%
over a neutrino path length $l\simeq t$.

In the simplest generalisation to a pure-to-mixed evolution, we have \cite%
{Lisi00}%
\begin{equation*}
\dot{P}_{x}=kP_{y}-\gamma P_{x}\;,\;\;\dot{P}_{y}=-kP_{x}-\gamma P_{y}\;,\;\;%
\dot{P}_{z}=0\ ,
\end{equation*}
where $\gamma\geq0$ is a phenomenological parameter. An initial pure state $%
\hat{\rho}(0)=\left\vert \nu_{\mu}\right\rangle \left\langle \nu_{\mu
}\right\vert $ now evolves into a mixed state with%
\begin{equation}
\mathbf{P}(t)=(e^{-\gamma t}\sin2\alpha\cos kt,-e^{-\gamma t}\sin2\alpha\sin
kt,\cos2\alpha)  \label{PG1}
\end{equation}
(where now $\left\vert \mathbf{P}(t)\right\vert <1$), and the oscillations
in the survival probability%
\begin{equation*}
p_{\mathrm{QT}}^{\mu}(t)=1-\frac{1}{2}(1-e^{-\gamma t}\cos kt)\sin^{2}2\alpha
\end{equation*}
are damped over distances $l\gtrsim1/\gamma$. The (initially zero) von
Neumann entropy $S_{\mathrm{vonN}}(t)$ increases with time, reaching a
limiting value%
\begin{equation*}
S_{\mathrm{vonN}}(\infty)=-\cos^{2}\alpha\ln(\cos^{2}\alpha)-\sin^{2}\alpha
\ln(\sin^{2}\alpha)\ .
\end{equation*}

If such pure-to-mixed transitions exist, it is possible that they are
accompanied by a transition from quantum equilibrium to quantum
nonequilibrium, along the lines considered in section 5. Applying the ansatz
(\ref{ConsS}), the nonequilibrium distribution would satisfy the constraint%
\begin{equation}
S_{\mathrm{hv}}(t)=-S_{\mathrm{vonN}}(t)\ ,  \label{Shvnu}
\end{equation}%
where in a general hidden-variables theory $S_{\mathrm{hv}}$ takes the form%
\begin{equation*}
S_{\mathrm{hv}}=-\int d\lambda \ \rho \ln (\rho /\rho _{\mathrm{QT}})\ .
\end{equation*}%
According to (\ref{Shvnu}), the hidden-variable entropy $S_{\mathrm{hv}}$
decreases with path length $l\simeq t$, in a manner that is fully determined
by the dynamics of the pure-to-mixed transition.

Quantum nonequilibrium $\rho (\lambda )\neq \rho _{\mathrm{QT}}(\lambda )$
would change the composition of a neutrino beam, in a manner depending on
the details of the hidden-variables theory. Generally speaking, the quantum
survival probability for $\nu _{\mu }$ may be written as%
\begin{equation}
p_{\mathrm{QT}}^{\mu }(t)=\frac{1}{2}(1+\mathbf{P}(0)\cdot \mathbf{P}(t))\ ,
\end{equation}%
which is again Malus' law (\ref{p+QT}) for a two-state system: $p_{\mathrm{QT%
}}^{\mu }(t)$ is just the probability $p_{\mathrm{QT}}^{+}(\mathbf{m})$ at
time $t$ for an `up' outcome of a quantum measurement along the axis
specified by the unit vector $\mathbf{m}=\mathbf{P}(0)$ in Bloch space
(corresponding to measuring for the presence of $\nu _{\mu }$), where the
measurement is carried out on a system with polarisation $\mathbf{P}(t)$. As
discussed in section 2, the probability law (\ref{p+QT}) is equivalent to
expectation additivity for incompatible observables, and both are
generically violated in nonequilibrium \cite{Sig}.

For example, applying the quadrupole probability law (\ref{QUAD}) to the
case at hand, we have a nonequilibrium survival probability for $\nu _{\mu }$%
,%
\begin{equation}
p^{\mu }(t)=p_{\mathrm{QT}}^{\mu }(t)+\frac{1}{2}\left( \mathbf{P}(0)\cdot 
\mathbf{b}(t)\right) \left( \mathbf{P}(0)\cdot \mathbf{P}(t)\right) \ ,
\end{equation}%
where $\mathbf{P}(0)$ and $\mathbf{P}(t)$ are given by (\ref{P0}) and (\ref%
{PG1}) respectively, and where the time dependence of $\mathbf{b}(t)$ (with $%
\mathbf{b}(0)=0$) corresponds to the generation of nonequilibrium during the
pure-to-mixed transition (perhaps in accordance with the constraint (\ref%
{Shvnu})). In the limit $t\rightarrow \infty $, for example, the composition
of the beam is shifted from the quantum $\nu _{\mu }$ fraction%
\begin{equation}
p_{\mathrm{QT}}^{\mu }(\infty )=1-\frac{1}{2}\sin ^{2}2\alpha
\end{equation}%
to the nonequilibrium $\nu _{\mu }$ fraction%
\begin{equation}
p^{\mu }(\infty )=p_{\mathrm{QT}}^{\mu }(\infty )+\frac{1}{2}\left(
b_{x}(\infty )\sin 2\alpha +b_{z}(\infty )\cos 2\alpha \right) \cos
^{2}2\alpha \ .
\end{equation}

\section{Particles from Very Distant Sources}

Finally, we consider a method for testing quantum probabilities at tiny
lengthscales, a method that is based on the huge spreading of the wave
packet for particles emitted by very distant (astrophysical or cosmological)
sources. In the right circumstances, such spreading can cause microscopic
deviations from the Born rule (if they exist) to be expanded up to
observable lengthscales. We shall restrict ourselves here to the case of
pilot-wave theory, though the argument can be generalised. As we shall see,
there are a number of practical difficulties with this method, and it is
unclear whether they could all be overcome in a real experiment. Still, the
idea might be worth considering further.

To explain the basic mechanism, we first consider a single nonrelativistic
particle (labelled $i$) in free space, with initial wave function $\psi _{i}(%
\mathbf{x},0)$ (at $t=0$) localised around $\mathbf{x}_{i}$ with a width $%
\Delta _{i}(0)$, where at later times $\psi _{i}(\mathbf{x},t)$ spreads out
to a width $\Delta _{i}(t)$. For large $t$, we have approximately $\Delta
_{i}(t)\sim \hslash t/\left( m\Delta _{i}(0)\right) $ (where $\sim \hslash
/\Delta _{i}(0)$ is the initial quantum momentum spread). One might think,
for example, of a spreading Gaussian packet. Now consider (in pilot-wave
theory) the time evolution of an initial distribution $\rho _{i}(\mathbf{x}%
,0)$ that differs from $\left\vert \psi _{i}(\mathbf{x},0)\right\vert ^{2}$
at a `nonequilibrium lengthscale' $\delta (0)$ at $t=0$. (We mean this in
the following sense: if $\rho _{i}(\mathbf{x},0)$ and $\left\vert \psi _{i}(%
\mathbf{x},0)\right\vert ^{2}$ are each coarse-grained or averaged over a
volume $\varepsilon ^{3}$, the difference between them is erased if and only
if $\varepsilon >>\delta (0)$.) Because $\rho _{i}$ and $\left\vert \psi
_{i}\right\vert ^{2}$ obey the same continuity equation, with the same (de
Broglian) velocity field, the ratio $f_{i}(\mathbf{x},t)\equiv \rho _{i}(%
\mathbf{x},t)/\left\vert \psi _{i}(\mathbf{x},t)\right\vert ^{2}$ is
conserved along particle trajectories (where nonequilibrium corresponds to $%
f_{i}\neq 1$). Thus, along a trajectory $\mathbf{x}(t)\equiv g_{t}(\mathbf{x}%
(0))$ we have $f_{i}(\mathbf{x}(t),t)=f_{i}(\mathbf{x}(0),0)$, and the
distribution at time $t$ may be written as%
\begin{equation}
\rho _{i}(\mathbf{x},t)=\left\vert \psi _{i}(\mathbf{x},t)\right\vert
^{2}f_{i}(g_{t}^{-1}(\mathbf{x}),0)\ ,  \label{f0}
\end{equation}%
where $g_{t}^{-1}$ is the inverse map from $\mathbf{x}(t)$ to $\mathbf{x}(0)$%
. If the map $g_{t}:\mathbf{x}(0)\rightarrow \mathbf{x}(t)$ is essentially
an expansion --- with small (localised) volumes $V_{0}$ of $\mathbf{x}(0)$%
-space being mapped to large volumes $V_{t}$ of $\mathbf{x}(t)$-space ---
then the inverse map $g_{t}^{-1}:\mathbf{x}(t)\rightarrow \mathbf{x}(0)$
will be essentially a compression. And because the spreading of $\left\vert
\psi _{i}\right\vert ^{2}$ is precisely the spreading of an initial
equilibrium distribution by the same map $g_{t}$, the factor by which $g_{t}$
expands an initial volume will be approximately $\sim \left( \Delta
_{i}(t)/\Delta _{i}(0)\right) ^{3}$, so that $V_{t}\sim \left( \Delta
_{i}(t)/\Delta _{i}(0)\right) ^{3}V_{0}$. Thus, if $f_{i}(\mathbf{x},0)$
deviates from unity on a lengthscale $\delta (0)$, then $f_{i}(g_{t}^{-1}(%
\mathbf{x}),0)$ will deviate from unity on an expanded lengthscale%
\begin{equation}
\delta (t)\sim \left( \Delta _{i}(t)/\Delta _{i}(0)\right) \delta (0)\ .
\label{deltat}
\end{equation}%
Therefore, from (\ref{f0}), the distribution $\rho _{i}(\mathbf{x},t)$ at
time $t$ will show deviations from $\left\vert \psi _{i}(\mathbf{x}%
,t)\right\vert ^{2}$ on the \textit{expanded} nonequilibrium lengthscale $%
\delta (t)$ \cite{AV96,AV01}.

As a simple example (assuming that the above nonrelativistic reasoning
extends to photons in some appropriate way), consider a photon with an
initial wave packet width $\Delta _{i}(0)\sim 10^{-6}\ \mathrm{cm}$, emitted
by an atom in the neighbourhood of a quasar at a distance $d\sim 10^{27}\ 
\mathrm{cm}$. The expansion factor is $\Delta _{i}(t)/\Delta _{i}(0)\sim
d/\Delta _{i}(0)\sim 10^{33}$, and an initial nonequilibrium lengthscale of
(for example) $\delta (0)\sim 10^{-33}$ $\mathrm{cm}$ is expanded up to $%
\delta (t)\sim 1$ $\mathrm{cm}$. (A photon would of course be found on the
surface of a sphere of radius $ct$, but distances on the spherical surface
still expand by a factor $\sim d/\Delta _{i}(0)$.)

So far we have considered the ideal case of a pure ensemble of identical
initial wave functions $\psi _{i}(\mathbf{x},0)$ centred around the same
point $\mathbf{x}_{i}$ and expanding in free space. To be realistic, we need
to consider a mixed ensemble emitted by a source of finite spatial extent%
\footnote{%
Averaging over the spatial extent of the source is important here because
the hidden variables are particle positions --- and not some more abstract
(or perhaps internal) degrees of freedom $\lambda $ whose distribution might
be independent of the spatial location of the emission.} and propagating in
a tenuous (intergalactic) medium.

Let the initial density operator be a mixture%
\begin{equation*}
\hat{\rho}(0)=\sum_{i}p_{i}\left\vert \psi _{i}\right\rangle \left\langle
\psi _{i}\right\vert
\end{equation*}%
of wave functions $\psi _{i}(\mathbf{x},0)$ centred at different points $%
\mathbf{x}_{i}$, with $p_{i}$ being the probability for the $i$th state. For
simplicity, let us first assume that $\psi _{i}(\mathbf{x},0)=\psi (\mathbf{x%
}-\mathbf{x}_{i},0)$, so that we have a mixture with the `same' wave
function spreading out from different locations $\mathbf{x}_{i}$. The
quantum equilibrium probability density at time $t$ is%
\begin{equation}
\rho _{\mathrm{QT}}(\mathbf{x},t)=\left\langle \mathbf{x}\right\vert \hat{%
\rho}(t)\left\vert \mathbf{x}\right\rangle =\sum_{i}p_{i}\left\vert \psi (%
\mathbf{x}-\mathbf{x}_{i},t)\right\vert ^{2}\ .  \label{rhoQT}
\end{equation}%
For each (quantum-theoretically) pure subensemble with guiding wave function 
$\psi _{i}(\mathbf{x},t)$, we may define an actual distribution $\rho _{i}(%
\mathbf{x},t)$ (generally distinct from $\left\vert \psi _{i}(\mathbf{x}%
,t)\right\vert ^{2}$), while for the whole ensemble the distribution may be
written as%
\begin{equation*}
\rho (\mathbf{x},t)=\sum_{i}p_{i}\rho _{i}(\mathbf{x},t)
\end{equation*}%
(where in general $\rho (\mathbf{x},t)\neq \rho _{\mathrm{QT}}(\mathbf{x},t)$%
). Let us also assume that, at $t=0$, each $\rho _{i}(\mathbf{x},0)$ takes
the form $\rho _{i}(\mathbf{x},0)=\pi (\mathbf{x}-\mathbf{x}_{i},0)$, where $%
\pi (\mathbf{x}-\mathbf{x}_{i},0)$ deviates from $\left\vert \psi (\mathbf{x}%
-\mathbf{x}_{i},0)\right\vert ^{2}$ at a nonequilibrium lengthscale $\delta
(0)$, so that we have a mixture with the `same' nonequilibrium distribution $%
\pi (\mathbf{x}-\mathbf{x}_{i},t)$ spreading out from different locations $%
\mathbf{x}_{i}$. (In work to be published elsewhere, we shall consider
dropping this last assumption.) The ensemble distribution is then%
\begin{equation}
\rho (\mathbf{x},t)=\sum_{i}p_{i}\pi (\mathbf{x}-\mathbf{x}_{i},t)\ .
\label{rho}
\end{equation}%
From our discussion of the pure case, we know that $\pi (\mathbf{x}-\mathbf{x%
}_{i},t)$ will deviate from $\left\vert \psi (\mathbf{x}-\mathbf{x}%
_{i},t)\right\vert ^{2}$ on an expanded lengthscale $\delta (t)\sim \left(
\Delta (t)/\Delta (0)\right) \delta (0)$, where $\Delta (t)$ is the width of 
$\left\vert \psi \right\vert ^{2}$ at time $t$.

Will a similar difference be visible between the spatially-averaged
distributions (\ref{rhoQT}) and (\ref{rho})? The answer depends on whether
the linear size $R$ of the source is larger or smaller than the (pure)
expanded nonequilibrium lengthscale $\delta (t)$. If $R>>\delta (t)$, the
spatial averaging will erase the differences between $\pi (\mathbf{x}-%
\mathbf{x}_{i},t)$ and $\left\vert \psi (\mathbf{x}-\mathbf{x}%
_{i},t)\right\vert ^{2}$, resulting in $\rho (\mathbf{x},t)\approx \rho _{%
\mathrm{QT}}(\mathbf{x},t)$. If, on the other hand, $R\lesssim \delta (t)$,
the spatial averaging cannot erase the nonequilibrium, and the observed
ensemble distribution $\rho (\mathbf{x},t)$ will deviate from the quantum
expression $\rho _{\mathrm{QT}}(\mathbf{x},t)$ on the expanded lengthscale $%
\delta (t)$.

We then arrive at the following conclusion (tentatively ignoring the effects
of scattering and of a mixture of different wave functions $\psi _{i}(%
\mathbf{x},0)\neq \psi (\mathbf{x}-\mathbf{x}_{i},0)$). For a distant source
of linear extension $R$, the spreading of wave packets from an initial width 
$\Delta (0)$ to a larger width $\Delta (t)$ will generate an observable
expansion of the nonequilibrium lengthscale from $\delta (0)$ to $\delta
(t)\sim \left( \Delta (t)/\Delta (0)\right) \delta (0)$, provided the `no
smearing' condition%
\begin{equation}
R\lesssim \delta (t)  \label{R}
\end{equation}%
is satisfied.

Before examining the feasibility of (\ref{R}) ever being satisfied in
practice, let us first indicate how our analysis --- carried out so far in
free space --- may be extended to include the effect of scattering by the
tenuous intergalactic medium.

Our strategy is as follows. We write the perturbed de Broglie-Bohm
trajectory $\mathbf{x}(t)$ (guided by a perturbed wave function that
includes scattering terms) as $\mathbf{x}(t)=\mathbf{x}_{\mathrm{free}%
}(t)+\delta \mathbf{x}(t)$, where $\mathbf{x}_{\mathrm{free}}(t)$ denotes
the trajectory in free space. As we have seen, the spreading of the
trajectories $\mathbf{x}_{\mathrm{free}}(t)$ generates an expanding
nonequilibrium lengthscale $\delta (t)$. The question is: will the
trajectory perturbations $\delta \mathbf{x}(t)$ cause the expanding
nonequilibrium to relax to equilibrium? Considering again the property of de
Broglian dynamics, that (for a pure subensemble) $f\equiv \rho /\left\vert
\psi \right\vert ^{2}$ is conserved along trajectories, a little thought
shows that a necessary condition for the erasure of nonequilibrium on the
expanded lengthscale $\delta (t)$ is that the perturbations $\delta \mathbf{x%
}(t)$ have a magnitude at least comparable to $\delta (t)$. If, on the
contrary,%
\begin{equation}
\left\vert \delta \mathbf{x}(t)\right\vert <<\delta (t)\ ,  \label{Scatt}
\end{equation}%
it will be impossible for the perturbations to erase the expanding
nonequilibrium --- simply because the trajectories will not be able to
distribute the values of $f$ in a manner required for the distributions $%
\rho $ and $\left\vert \psi \right\vert ^{2}$ to become indistinguishable on
a coarse-graining scale of order $\delta (t)$.

A straightforward argument suggests that the `no relaxation' condition (\ref%
{Scatt}) is indeed likely to be realised in practice. To estimate the
magnitude $\left\vert \delta \mathbf{x}(t)\right\vert $, at large distances
from the source we may approximate the wave function as a plane wave $e^{i%
\mathbf{k}\cdot \mathbf{x}}$ incident on a tenuous medium modelled by fixed
scattering centres with positions $\mathbf{x}_{s}$. In a time-independent
description of the scattering process, each scattering centre (associated
with some potential) contributes a scattered wave which, at large distances
from $\mathbf{x}_{s}$, takes the asymptotic form $f_{s}(\theta ,\phi
)e^{ikr_{s}}/r_{s}$, where $r_{s}\equiv \left\vert \mathbf{x}-\mathbf{x}%
_{s}\right\vert $ and $(\theta ,\phi )$ are standard angular coordinates
defined relative to $\mathbf{k}$ as the `$z$-axis'. The scattering amplitude 
$f_{s}(\theta ,\phi )$ is related to the differential cross section by the
usual formula $d\sigma _{s}/d\Omega =\left\vert f_{s}(\theta ,\phi
)\right\vert ^{2}$. For simplicity we may consider identical and isotropic
scattering centres: $f_{s}(\theta ,\phi )=f=\mathrm{const}.$ for all $s$, so
that $f^{2}=\sigma /4\pi $ where $\sigma $ is the cross section. The total
(time-independent) wave function is then%
\begin{equation}
\psi (\mathbf{x})=e^{i\mathbf{k}\cdot \mathbf{x}}-\frac{1}{2}\sqrt{\frac{%
\sigma }{\pi }}\sum_{s}e^{i\mathbf{k}\cdot \mathbf{x}_{s}}\frac{e^{ikr_{s}}}{%
r_{s}}  \label{Psi}
\end{equation}%
(where $e^{i\mathbf{k}\cdot \mathbf{x}_{s}}$ is a relative phase for each
source). We assume that the scattering centres are more or less uniformly
distributed in space with a number density $n$ and mean spacing $(1/n)^{1/3}$%
. The intergalactic medium is mainly composed of ionised hydrogen, with an
electron number density $n\sim 10^{-7}$\ $\mathrm{cm}^{-3}$ and mean spacing 
$\sim 200\ \mathrm{cm}$. For most cases of interest, the incident wavelength 
$2\pi /k$ will be much smaller than $200\ \mathrm{cm}$, justifying use of
the asymptotic form $\sim e^{ikr_{s}}/r_{s}$ for the scattered waves. (In an
appropriate extension of this nonrelativistic model to photons, wavelengths $%
2\pi /k\gtrsim 200\ \mathrm{cm}$ correspond to radio waves.) In this
approximation, the de Broglie-Bohm trajectories take the form%
\begin{equation*}
\mathbf{x}(t)=\mathbf{x}(0)+(\hslash \mathbf{k}/m)t+\delta \mathbf{x}(t)\ ,
\end{equation*}%
where $\delta \mathbf{x}(t)$ is a small perturbation. We expect $\delta 
\mathbf{x}(t)$ to behave like a random walk, with $\left\vert \delta \mathbf{%
x}(t)\right\vert \propto \sqrt{t}$. If this is the case, then because $%
\delta (t)\propto \Delta (t)\propto t$ (for large $t$), the no relaxation
condition (\ref{Scatt}) will necessarily be satisfied for sufficiently large 
$t$. It then appears that scattering by the intergalactic medium is unlikely
to offset the expansion of the nonequilibrium lengthscale.

In contrast, the no smearing condition (\ref{R}) is very severe, and is
unlikely to be realised except in special circumstances. In our example
above, of a photon emitted by an atom in the vicinity of a quasar, the
expansion factor $\Delta (t)/\Delta (0)\sim 10^{33}$ suggests that the
tantalising Planck lengthscale $l_{\mathrm{P}}\sim 10^{-33}$ $\mathrm{cm}$
at the time of emission may be within reach of experiments performed on the
detected photon now at a lengthscale $\sim 1$ $\mathrm{cm}$. Unfortunately,
according to (\ref{R}) any nonequilibrium at the Planck scale would be
smeared out unless the source had a size $R\lesssim 1$ $\mathrm{cm}$, which
seems much too small to be resolvable in practice (at the assumed distance $%
d\sim 10^{27}$ $\mathrm{cm}$).

This seemingly insurmountable obstacle could perhaps be overcome, however,
by considering a combination of: (a) shorter wavelengths, corresponding to a
smaller $\Delta (0)$ and a larger $\delta (t)$; and (b) special
astrophysical circumstances in which remarkably small sources can in fact be
resolved.

As an example of (a), one may consider a gamma-ray emission (say from an
atomic nucleus) with $\Delta(0)\sim10^{-12}$ $\mathrm{cm}$, again from a
distance $d\sim10^{27}$ $\mathrm{cm}$, yielding an expansion factor $%
\Delta(t)/\Delta(0)\sim d/\Delta(0)\sim10^{39}$. To probe the Planck scale ($%
\delta(0)\sim10^{-33}$ $\mathrm{cm}$) then requires a source size $%
R\lesssim\delta(t)\sim10^{6}$ $\mathrm{cm}=10$ $\mathrm{km}$, which is
comparable to what is believed to be the size of the central engine of a
typical gamma-ray burst \cite{Mes06}. (Photons from the central engine of a
gamma-ray burst are not normally expected to propagate essentially freely
immediately after emission, and there are in any case many uncertainties
concerning the mechanism of such bursts; even so, the example just quoted
does suggest that the no smearing constraint (\ref{R}) might in fact be
satisfied by a judicious choice of wavelength and source.)

As examples of (b), we quote the following instances of remarkably small
sources that either have already been resolved in practice, or that might be
in the near future:

\begin{itemize}
\item Nanosecond radio bursts have been observed coming from the Crab pulsar 
\cite{Han03}. The observations have a time resolution $\Delta t\approx 2\ 
\mathrm{ns}$, corresponding to an emitting source diameter $\lesssim c\Delta
t\approx 60\ \mathrm{cm}$. Isolated sub-pulses were detected at this time
resolution, and interpreted as caused by the collapse of highly-localised ($%
\sim 60\ \mathrm{cm}$) structures in a turbulent plasma. Whatever their
nature, these objects are the smallest ever resolved outside the solar
system. For our purposes, the Crab pulsar is too close ($d\sim 10^{22}\ 
\mathrm{cm}$) and radio wavelengths are too large. Even so, it is clear that
the detection of transients on very small timescales --- at an appropriate
distance and wavelength --- offers a way of resolving sources satisfying the
no smearing condition (\ref{R}).

\item An ultraviolet ($\approx170\ \mathrm{eV}$) `hotspot' of radius $%
\lesssim60\ \mathrm{m}$ has been detected on the surface of the Geminga
pulsar (at a distance $d\sim10^{21}\ \mathrm{cm}$ from Earth) \cite{Car04}.
While the source is again too close for our purposes, both the wavelength
and the source size are promising.

\item Microsecond gamma-ray bursts of energy $\gtrsim 100\ \mathrm{MeV}$ (or
wavelength $\lesssim 10^{-12}\ \mathrm{cm}$) --- which might originate from
exploding primordial black holes --- should be observable with the SGARFACE
experiment \cite{LeB05}. A burst time structure with resolution $\Delta
t\approx 10^{-6}\ \mathrm{s}$ corresponds to a source size $\lesssim c\Delta
t\approx 10^{4}\ \mathrm{cm}=0.1\ \mathrm{km}$. Microsecond gamma-ray bursts
at cosmological distances ($d\sim 10^{27}-10^{28}\ \mathrm{cm}$) would then
seem to satisfy our criteria (except that the possibility of essentially
free propagation from emission to detection still needs to be considered as
well).
\end{itemize}

Finally, we must consider dropping the simplifying assumption that the
packets $\psi _{i}(\mathbf{x},0)$ emitted by the source differ only in their
initial location. In general, we will have $\psi _{i}(\mathbf{x},0)\neq \psi
(\mathbf{x}-\mathbf{x}_{i},0)$, with different packets $\psi _{i}$ emitted
from different locations $\mathbf{x}_{i}$. Here we seem to encounter the
most severe practical problem of all. As in the case of perturbations from
the intergalactic medium, a necessary condition for the different wave
functions to lead to an erasure of nonequilibrium on the expanded
lengthscale $\delta (t)$ is that trajectories $\mathbf{x}_{i}(t)$, $\mathbf{x%
}_{j}(t)$ (with the same initial point $\mathbf{x}(0)$) generated by
respective wave functions $\psi _{i}(\mathbf{x},t)$, $\psi _{j}(\mathbf{x}%
,t) $ should differ by an amount at least comparable to $\delta (t)$. If
instead%
\begin{equation}
\left\vert \mathbf{x}_{i}(t)-\mathbf{x}_{j}(t)\right\vert \lesssim \delta (t)
\label{mix}
\end{equation}%
for all $i$, $j$ (and for all $\mathbf{x}(0)$), it will be impossible for
the expanding nonequilibrium to be erased upon averaging over the mixture of
wave functions.

Unfortunately, it is unclear whether the `no mixing' condition (\ref{mix})
could ever be realised in practice. The wave functions $\psi _{i}(\mathbf{x}%
,t)$, $\psi _{j}(\mathbf{x},t)$ would have to be almost the same, to an
extremely high accuracy, in order to generate trajectories $\mathbf{x}%
_{i}(t) $, $\mathbf{x}_{j}(t)$ satisfying (\ref{mix}). To see this, as a
rough estimate one may take $\mathbf{x}_{i}(t)\sim (\Delta \mathbf{p}%
_{i}/m)t $, where $\Delta \mathbf{p}_{i}$ is the quantum momentum spread for
the wave function $\psi _{i}$; and similarly for $\mathbf{x}_{j}(t)$. The
condition (\ref{mix}) then reads%
\begin{equation}
\left\vert \Delta \mathbf{p}_{i}-\Delta \mathbf{p}_{j}\right\vert \lesssim 
\frac{\hslash }{\Delta (0)}\frac{\delta (0)}{\Delta (0)}  \label{mom}
\end{equation}%
(taking all initial packets to have approximately the same width $\Delta (0)$
and nonequilibrium lengthscale $\delta (0)$). Since the factor $\delta
(0)/\Delta (0)$ is very tiny, the momentum spreads of the emitted packets
must be very tightly constrained, and there seems to be no obvious way in
which this could happen.

If the method proposed in this section is to work in practice, some extra
ingredient is needed to ensure that (\ref{mom}) is satisfied. At the time of
writing, we are unable to say if such an ingredient is likely to be found.

\section{Conclusion}

We have discussed several proposals for astrophysical and cosmological tests
of quantum theory. Our general aim has been to test the foundations of
quantum theory in new and extreme conditions, guided in particular by the
view that quantum theory is an emergent description of an equilibrium state.
While we have often used the pilot-wave theory of de Broglie and Bohm, much
of our reasoning applies to general deterministic hidden-variables theories.

Pilot-wave theory is the only hidden-variables theory of broad scope that we
possess. Possibly, it is a good approximation to the correct theory; or
perhaps it is merely a helpful stepping stone towards the correct theory.
Certainly, pilot-wave theory is a simple and natural deterministic
interpretation of quantum physics. On the other hand, it could be that the
true deterministic hidden-variables theory (if there is one) is quite
different, and that in some key respects pilot-wave theory is actually
misleading. After all, the observable statistics of the quantum equilibrium
state obscure many of the details of the underlying (nonlocal and
deterministic) physics. Since all of our experience so far has been confined
to the equilibrium state, it would not be surprising if we were led astray
in our attempts to construct a subquantum (or hidden-variables) theory.
Obviously, many possible theories could underlie the equilibrium physics
that we see. The ultimate aim of the proposals made in this paper is to find
an empirical window that could help us determine what the true underlying
theory actually is.

It is usually assumed that quantum theory is a fundamental framework in
terms of which all physical theories are to be expressed. There is, however,
no reason to believe \textit{a priori} that quantum theory has an unlimited
domain of validity. For two hundred years it was generally believed that
Newtonian mechanics was a fundamental framework for the whole of physics.
Yet, today we know that Newtonian mechanics is merely an emergent
approximation (arising from the classical and low-energy limits of quantum
field theory). Whether or not quantum theory will suffer a similar fate
remains to be seen.


\begin{thebibliography}{99}
\bibitem{P76} P. Pearle, Phys. Rev. D \textbf{13}, 857 (1976).

\bibitem{P79} P. Pearle, Int. J. Theor. Phys. \textbf{18}, 489 (1979).

\bibitem{P89} P. Pearle, Phys. Rev. A \textbf{39}, 2277 (1989).

\bibitem{GRW86} G.-C. Ghirardi, A. Rimini and T. Weber, Phys. Rev. D \textbf{%
34}, 470 (1986).

\bibitem{AV91a} A. Valentini, Phys. Lett. A \textbf{156}, 5 (1991).

\bibitem{AV91b} A. Valentini, Phys. Lett. A \textbf{158}, 1 (1991).

\bibitem{AV92} A. Valentini, PhD thesis, International School for Advanced
Studies, Trieste, Italy (1992) [www.sissa.it/ap/PhD/Theses/valentini.pdf].

\bibitem{AV96} A. Valentini, in: \textit{Bohmian Mechanics and Quantum
Theory: an Appraisal}, eds. J.T. Cushing, A. Fine, and S. Goldstein (Kluwer,
Dordrecht, 1996).

\bibitem{AV01} A. Valentini, in: \textit{Chance in Physics: Foundations and
Perspectives}, eds. J. Bricmont \textit{et al}. (Springer, Berlin, 2001).

\bibitem{AV02} A. Valentini, Phys. Lett. A \textbf{297}, 273 (2002).

\bibitem{AVPr02} A. Valentini, Pramana -- J. Phys. \textbf{59}, 269 (2002).

\bibitem{Sig} A. Valentini, Phys. Lett. A \textbf{332}, 187 (2004).

\bibitem{PV06} P. Pearle and A. Valentini, in: \textit{Encyclopaedia of
Mathematical Physics}, eds. J.-P. Fran\c{c}oise, G. Naber and T. S. Tsun
(Elsevier, 2006).

\bibitem{deB28} L. de Broglie, in: \textit{\'{E}lectrons et Photons:
Rapports et Discussions du Cinqui\`{e}me Conseil de Physique}, ed. J. Bordet
(Gauthier-Villars, Paris, 1928). [English translation in ref. \cite{BV06}.]

\bibitem{B52} D. Bohm, Phys. Rev. \textbf{85}, 166; 180 (1952).

\bibitem{BV06} G. Bacciagaluppi and A. Valentini, \textit{Quantum Theory at
the Crossroads: Reconsidering the 1927 Solvay Conference} (Cambridge
University Press, forthcoming) [quant-ph/0609184].

\bibitem{VW05} A. Valentini and H. Westman, Proc. Roy. Soc. Lond. A \textbf{%
461}, 253 (2005).

\bibitem{Sky} P. Fernstr\"{o}m, J. Johansson and A. Skyman, Chalmers
University of Technology unpublished report.

\bibitem{AV06} A. Valentini, Inflationary cosmology as a probe of primordial
quantum mechanics, in preparation.

\bibitem{AV04} A. Valentini, hep-th/0407032.

\bibitem{LL} A. R. Liddle and D. H. Lyth, \textit{Cosmological Inflation and
Large-Scale Structure} (Cambridge University Press, 2000).

\bibitem{CMBdata} G. Hinshaw \textit{et al}., astro-ph/0603451.

\bibitem{PSS06} A. Perez, H. Sahlmann and D. Sudarsky, Class. Quantum Grav. 
\textbf{23}, 2317 (2006).

\bibitem{H76} S. W. Hawking, Phys. Rev. D \textbf{14}, 2460 (1976).

\bibitem{Page} D.N. Page, Phys. Rev. D \textbf{13}, 198 (1976); Phys. Rev. D 
\textbf{14}, 3260 (1976); Phys. Rev. D \textbf{16}, 2402 (1977).

\bibitem{AHDD} N. Arkani-Hamed, S. Dimopoulos and G. Dvali, Phys. Lett. B 
\textbf{429}, 263 (1998).

\bibitem{Dan93} U. H. Danielsson and M. Schiffer, Phys. Rev. D \textbf{48},
4779 (1993).

\bibitem{FF05} L. Ferrarese and H. Ford, Space Sci. Rev. \textbf{116}, 523
(2005).

\bibitem{RN03} C. S. Reynolds and M. A. Nowak, Phys. Rep. \textbf{377}, 389
(2003).

\bibitem{AV04b} A. Valentini, astro-ph/0412503.

\bibitem{Asp02} A. Aspect, in: \textit{Quantum (Un)speakables: From Bell to
Quantum Information}, eds. R. A. Bertlmann and A. Zeilinger (Springer,
Berlin, 2002).

\bibitem{C75} C. T. Cunningham, Astrophys. J. \textbf{202}, 788 (1975).

\bibitem{Ogle04} P. M. Ogle \textit{et al}., Astrophys. J. \textbf{606}, 151
(2004).

\bibitem{Lisi00} E. Lisi, A. Marrone and D. Montanino, Phys. Rev. Lett.%
\textit{\ }\textbf{85}, 1166 (2000).

\bibitem{Mav04} N. E. Mavromatos, gr-qc/0407005.

\bibitem{Anc05} L. A. Anchordoqui \textit{et al}., Phys. Rev. D \textbf{72},
065019 (2005).

\bibitem{Hoop05} D. Hooper, D. Morgan and E. Winstanley, Phys. Lett. B 
\textbf{609}, 206 (2005).

\bibitem{Chr05} J. Christian, Phys. Rev. Lett. \textbf{95}, 160403 (2005).

\bibitem{Mor06} D. Morgan, E. Winstanley, J. Brunner and L. F. Thompson,
Astropart. Phys. \textbf{25}, 311 (2006).

\bibitem{Bar06} G. Barenboim, N. E. Mavromatos, S. Sarkar and A.
Waldron-Lauda, hep-ph/0603028.

\bibitem{Beu03} M. Beuthe, Phys. Rep. \textbf{375}, 105 (2003).

\bibitem{Mes06} P. M\'{e}sz\'{a}ros, Rep. Prog. Phys. \textbf{69}, 2259
(2006).

\bibitem{Han03} T. H. Hankins, J. S. Kern, J. C. Weatherall and J. A. Eilek,
Nature \textbf{422}, 141 (2003).

\bibitem{Car04} P. A. Caraveo \textit{et al}., Science \textbf{305}, 376
(2004).

\bibitem{LeB05} S. LeBohec, F. Krennrich and G. Sleege, Astropart. Phys. 
\textbf{23}, 235 (2005).
\end{thebibliography}
\end{document}